# Photometric study of single-shot energy-dispersive X-ray diffraction at a laser plasma facility

O.R. Hoidn and G.T. Seidler[(*)]

Physics Department

University of Washington

Seattle WA 98195-1560

## Abstract

The low repetition rates and possible shot-to-shot variations in laser-plasma studies place a high value on single-shot diagnostics.  For example, white-beam scattering methods based on broadband backlighter x-ray sources are used to determine changes in the structure of laser-shocked crystalline materials by the evolution of coincidences of reciprocal lattice vectors and kinematically-allowed momentum transfers.  Here, we demonstrate that white-beam techniques can be extended to strongly-disordered dense plasma and warm dense matter (WDM) systems where reciprocal space is only weakly structured and spectroscopic detection is consequently needed to determine the static structure factor and thus the ion-ion radial distribution function.   Specifically, we report a photometric study of energy-dispersive diffraction (ED-XRD) for structural measurement of high energy density systems at large-scale laser facilities such as OMEGA and the National Ignition Facility.  We find that structural information can be obtained in single-shot ED-XRD experiments using established backlighter and spectrometer technologies.

(*) Corresponding author: seidler@uw.edu

# I Introduction

In addition to their centrality for inertial confinement fusion studies,[1,2] laser-shock experiments play a growing role at the interface between plasma physics and condensed matter physics, geosciences, and laboratory astrophysics.[3-14] However, for experiments reaching the highest energy density states the technical challenges extend beyond the creation of such states: the low repetition rates, limited facility access, and significant shot-to-shot variations each place a special emphasis on single-shot x-ray diagnostics of the structural and electronic properties of the compressed, heated target.[15-24] An important case-in-point is provided by the determination of the ion-ion radial distribution function, $g_{ii}(\vec{r})$, or equivalently the static structure factor $S(\vec{k})$. Knowledge of $g_{ii}(\vec{r})$ fulfills an interesting variety of roles. First, it is necessary, if only at the level of mean density and average ionization state, for investigation of any equations of state (EOS) and of molecular dynamics simulations or other structural calculations performed in support of EOS calculations. Second, it is also a critical input parameter to any fine treatment of electronic structure. The electronic structure of dense crystalline systems and plasmas, in turn, is a quantity of fundamental interest but also of a certain pragmatic interest: some sufficient knowledge of electronic structure is needed for reliable determination of the target temperature and ionization state in dense plasma and warm dense matter (WDM) experiments [25,26], and this capability is in turn needed for campaigns to experimentally measure the EOS in the WDM regime [27-29].

For targets that retain substantial medium- or long-range order upon shock compression, broadband backlighter x-ray sources enable white-beam angle-dispersive x-ray diffraction (AD-XRD) in which substantial structural detail can be inferred from Kossel rings [30] and other fine scattering patterns dictated by the coincidence of reciprocal lattice vectors and kinematically-allowed momentum transfers [31,32]. However, white-beam AD-XRD is only applicable to systems that are substantially single crystalline: any statistically isotropic system, whether a polycrystalline fine-powder sample or a dense, partially ionized plasma, when illuminated by a broad-band source will show an angularly-featureless signal when observed on, *e.g.*, an image plate. For high atomic number (Z) systems, single-shot white-beam extended x-ray absorption fine structure (EXAFS) has seen some applications [33]; the situation has proven more challenging for lower-Z WDM and dense plasmas, as a result of the mutually-exclusive target thickness requirements of the x-ray measurement (soft x-ray penetration lengths of order 1 micron or less) and laser ablation (necessary thicknesses of tens of microns) [34]. Consequently, the first determination of $g_{ii}(r)$ for disordered, dense lower-Z plasma systems [35] instead used multi-shot, quasi-monochromatic AD-XRD, *i.e.*, 'traditional' XRD.

Here, we investigate whether single-shot, white-beam XRD can be performed on *strongly disordered*, laser-shocked solids and WDM using spectral information at the detector location to parameterize the momentum transfer of the quasielastic scattering event, *i.e.*, we consider purely energy-dispersive x-ray diffraction (ED-XRD). Some context is needed to fully define this term and to distinguish it from XRD methods already in use in the laser-plasma community. The differential scattering cross section per atom for coherent scattering of x rays (ordinary diffraction of incoherent incident photons) from an isotropically disordered, elemental material such as a powder sample, liquid, or dense laser-shock heated plasma of a single atomic species is

$$\frac{d\sigma_{coh}}{d\Omega}(k) = \sigma_t S(k) f(k)^2 \qquad (1)$$

where $\sigma_t$ is the Thomson cross section, $S(k)$ is the directionally-averaged structure factor and $f(k)$ is the spherically averaged atomic form factor. The structure factor $S(k)$ is simply related by a sine transform to $g_{ii}(r)$,

$$S(k) = 1 + (4\pi\rho/k) \int_0^\infty dr\, r\, [\, g_{ii}(r) - 1\,]\, \sin(kr). \qquad (2)$$

These well-known expressions establish the close connection between XRD and $g_{ii}(r)$ while also demonstrating the need to measure the differential scattering cross-section (and hence $S(k)$) at many different momentum transfers if any significant constraint on the form of $g_{ii}(r)$ is to be obtained.

The $k$-dependence of $d\sigma_{coh}/d\Omega$ can in principle be measured with any suitable combinations of scattering angle $2\theta$ and photon energies spanning the needed momentum transfers: $k$ is chosen by the combined effect of these two experimentally-selectable parameters, $k = (2E/\hbar c)\sin 2\theta$. In practice, however, XRD is measured in only two modes: angle-resolved XRD (henceforth 'AR-XRD') and energy-dispersive XRD (henceforth 'ED-XRD'). Their distinction is best introduced kinematically. As illustrated in Fig. 1, any measurement of $S(k)$ must follow a curve in $E$-$2\theta$ space which crosses many of the shown contours of constant $k$. The parameter space probed by a typical AR-XRD experiment using ~8 keV monochromatic incident photons is represented by the vertical curve in the figure. Experimentally, the necessary apparatus will include a monochromatic source and either an angle-scanning detector or a position sensitive detector (PSD), which we show schematically in Fig. 2 (a) and (b). On the other hand, a typical ED-XRD experiment instead resides on the horizontal curve in Fig. 1, *i.e.*, at a fixed scattering angle of 135 degrees but requiring both a broad incident source spectrum and an energy-resolving detector. An experimental schematic for ED-XRD is presented in Fig. 2(c). We note that ED-XRD has a long history in laboratory and synchrotron XRD studies, and plays an important role in high-pressure diamond anvil cell research where the limited angular access to the sample space substantially complicates AD-XRD.[36-40] There is then an obvious commonality with experiments at large-scale laser facilities; angular access at such facilities is strongly constrained by the beam paths of the laser light itself.

AD-XRD from laser-shock compressed, disordered Al has recently been reported by Ma, *et al.*,[35] and this first such study illustrates both the scientific benefits and technical drawbacks of AD-XRD for large-scale laser facilities. Specifically, concerning the latter, a few high-resolution spectrometers must be moved between different scattering angles for different shots so as to obtain a complete characterization of $S(k)$ by pooling the results of many shots after suitable normalization or other characterization of shot-to-shot variations in the source or target. While the study of Ma, *et al.*,[35] has overcome these challenges and provides an interesting comparison of experiment to modern theoretical treatments of the structure of dense plasmas, it is still important to note that the use of a multi-shot technique has, at a minimum, decreased the range of phase space that can be studied subject to the strong constraints that exist on facility access. A single-shot alternative could therefore have high scientific impact and is likely the only way that $S(k)$ will be measured on disordered dense plasmas at the National Ignition Facility, where the number of shots per scientific study is especially limited.

Consequently, with the above context established, we report here a photometric analysis of ED-XRD for laser-shock experiments illuminated by broad-band backlighter sources. This analysis makes use of known results for the spectrum of a broad-band backlighter, representative experimental results for $S(k)$ for disordered systems, and representative, established technical characteristics of spectrally-resolving detectors available at large-scale laser facilities. We find significant benefits to ED-XRD for disordered systems, including single-shot determination of $S(k)$, and we propose that ED-XRD should become a standard diagnostic at large-scale facilities such as OMEGA and the National Ignition Facility

We continue as follows. In section 2 we describe the methods used in the photometric analysis, including the reference target, modeled experimental geometry, and any assumptions about detector or spectrometer performance. In section 3 we present and discuss our results for ED-XRD using each of two different experimental configurations. These are, first, an x-ray CCD detector operating in single-photon mode as an energy-resolving solid-state detector and, second, a wavelength-dispersive spectrometer using a highly-oriented pyrolytic graphite (HOPG) mosaic crystal as the diffractive element. The CCD configuration is viable, but has some drawbacks associated with saturation and double-counting that require special care. We find that the HOPG-based spectrometer quite easily resolves the energy spectrum of the diffraction with excellent counting statistics for a broad-band backlighter that has been fielded at OMEGA, with the caveat that a single HOPG crystal analyzer covers a narrower energy range, and hence a more restricted $k$-range, than a CCD detector. Finally, in section IV we conclude.

## II Methods

### II.A. Source and target

One readily available broadband source in laser shock experiments is the thermal spectrum from a laser-imploded polymer shell, usually filled with $H_2$-$D_2$ gas [41, 42]. In Fig. 3 we show a typical spectrum collected at OMEGA [43]. Because of the spectrum's supra-exponential decay, an ED-XRD experiment with this source is preferentially conducted at low energy, between 2 and 6 keV, as shown in the ED-XRD curve at a scattering angle of 135 degrees in Fig. 1. Also shown in Fig. 3 is the spectrum for a typical narrow band backlighter source at OMEGA, where these sources have seen extensive use in x-ray scattering studies, both elastic (XRD) [35] and inelastic (usually called 'x-ray Thomson scattering') [25]. The narrow-band spectrum is obtained by scaling the spectrum of a Cu $K_\alpha$ target driven by a 10 J, 10 ps laser pulse at the MTW laser facility to a 2.5 kJ, 10 ps laser pulse at OMEGA, using a typical $K_\alpha$ photon yield of $4 \times 10^{10}$ photons per J of laser energy [44, 45].

We consider two target systems where experimental $S(k)$ are available: liquid boron at ambient pressure and shock-compressed aluminum. For liquid boron we use the experimental results of Krishnan *et al.* [46, 47], the data for which were taken at a synchrotron light source using hydrodynamically-levitated boron heated to 2400K by continuous illumination from infrared lasers. While this is not a WDM system *per se*, it is a reasonable surrogate. As shown in Fig. 4 (a), note the presence of a few broad peaks in $S(k)$, representative of a system with only limited, short-range information in $g_{ii}(r)$. For clarity in our photometric analysis, we will use a smoothed $S(k)$ where the sharp (nonphysical) noise in the experimental $S(k)$ has been filtered. On the other hand, $S(k)$ for shock-compressed aluminum ($n_e$= 5.4 × $10^{23}$ cm$^{-3}$; $T_e$= 10 eV) is based on results from Ma *et al.*[35], who have recently reported the first AD-XRD measurement of a shock-compressed, disordered WDM system. $S(k)$ was recovered from Ma *et al.*'s theoretical calculation of an elastic scattering profile for triply-ionized shock-compressed aluminum, to which they fit their data. We note that only an approximate atomic form factor, that of ambient aluminum, was used to calculate $S(k)$ from the scattering profile; however, the resulting error in $S(k)$ is expected to be negligible above $k = 3\text{Å}^{-1}$, and hence does not affect the location of any coordination peaks. As shown in Fig. 4 (b), note again that the presence of only short-range order in the target results in a simple form for $S(k)$. In this case, the information content is largely limited to the location and intensity of the obvious first coordination peak.

**II.B. Photon-electron interactions and numerical modeling**

For the targets considered here, the experiment is conducted in an energy region far above any atomic fluorescence from the targets and also far above any soft x-ray blackbody radiation from the surface or bulk of the target, each of which is easily attenuated in practice with a thin plastic or Be shield. Consequently, we need only consider the coherent and incoherent scattering of the x-rays as direct contributors to the measured scattering signal; the photoelectric interaction appears only in its contribution to absorption coefficient in the energy range of interest. Note that by coherent here we refer to the quasielastic scattering process itself, *i.e.* "ordinary" diffraction, with no expectation of coherence of the incident beam (such as is used in diffraction experiments at XFEL facilities).

Given a backlighter source with fluence $I_{source}(E)$ (units of photons/eV, integrated over $4\pi$ steradian) at a distance $d_{source}$ from the target, the areal flux incident in the target is $I_{incident}(E) = I_{source}(E)/4\pi d_{source}^2$. The contribution of coherent scattering to the measured energy spectrum at a scattering angle $2\theta$ is then

$$I_{coh}(E, 2\theta) = I_{incident}(E) \frac{d\sigma_{coh}}{d\Omega}(k) \, d\Omega_{det} \, \eta_{det}(E) \, \tau_{coh}(E, 2\theta), \quad (3)$$

where $k$ is implicitly determined by $E$ and $2\theta$, $d\Omega_{det}$ is the solid angle subtended by the detector, $\eta_{det}(E)$ is the net efficiency of detection of photons of energy $E$ that arrive in $d\Omega_{det}$, and $\tau_{coh}(E, 2\theta)$ includes the necessary corrections to the measured XRD due to the target's geometry and energy-dependent absorption coefficient [34]. When operating near to a backscattering geometry, for example, $\tau_{sample}(E, 2\theta) \sim \rho A (1 - e^{-2\mu(E)d})/2\mu(E)$ where $\rho$ is atomic (number) density, $A$ is the cross-sectional area of the portion of the backlighter beam that illuminates the target region of interest, $d$ is the target thickness, and $\mu(E)$ is the x-ray absorption coefficient. For present purposes, $d\sigma_{coh}/d\Omega$ includes all elastic and quasielastic scattering; it integrates over all ion-ion correlation dynamics [48].

The incoherent contribution to the measured signal is somewhat more complex to model. The microscopic physics of the incoherent scattering processes, wherein one must address both momentum transfer ($k$) and energy transfer ($\hbar\omega$), results in the need for a double differential cross-section $d^2\sigma_{incoh}(k, \hbar\omega)/d\omega d\Omega$. The detected intensity from incoherent scattering is then

$$I_{incoh}(E, 2\theta) = d\Omega_{det} N_{atoms} \frac{d\sigma_t}{d\Omega} \int_0^\infty dE' I_{incident}(E') \, S_{incoh}(k, \omega) \tau_{incoh}(E', E, 2\theta) \quad (4),$$

where $N_{atoms}$ is the number of atoms in the target, $k$ is again implicitly determined by $E$, $E'$, and $2\theta$ and $\tau_{incoh}(E', E, 2\theta)$ includes the influence of attenuation for an incident photon of energy $E'$ that scatters through an angle $2\theta$ and departs the incoherent interaction with energy $E$. In the first Born approximation, $d^2\sigma_{incoh}/d\omega d\Omega = (d\sigma_t/d\Omega) S_{incoh}(k,\omega)$, where $d\sigma_t/d\Omega$ is the Thomson differential scattering cross section and $S_{incoh}(k,\omega)$ is the inelastic component of the dynamic structure factor. In the independent-electron approximation [49, 50], $S_{incoh}(k,\omega)$ may be expressed as a sum over electrons and matrix elements between the initial and final states of the system:

$$S_{incoh}(k,\omega) = \sum_j \sum_{f \neq i} |\langle i | e^{i\,q \cdot r_j} | f \rangle|^2 \delta(E_f - E_i - \hbar\omega). \quad (5)$$

At sufficiently high $k$, $S_{incoh}(k,\omega)$ is peaked at the Compton shift $\Delta E = \hbar^2 k^2 / 2m$. In the high-$k$ (non-collective) scattering regime the total inelastic portion of the dynamic structure factor is constructed using equation (5) evaluated as a sum over individual valence and core electrons. In our modeling procedure this consists of truncated valence and core Compton profiles generated in the impulse approximation [50, 51] where $S_{incoh}(k,\omega)$ depends only on the ground state electronic density and kinematics of the scattering process. The first moment of $S_{incoh}(k,\omega)$ was normalized after truncation according to the Bethe $f$-sum rule [52]. For boron our approximation yields incoherent scattering cross sections which exceed experimental values by up to 30 percent, making our approximate treatment of the incoherent background conservative. In the relevant range of momentum transfers the Compton shift is sufficiently small that we substitute $\tau_{coh}(E, 2\theta)$ for $\tau_{incoh}(E, E', 2\theta)$ without introducing appreciable systematic error, allowing $\tau$ to be factored out of the integrand in equation (4). Similarly, the FWHM of $S_{incoh}(k,\omega)$ (which, in the impulse approximation, is directly related to the width of the momentum distribution of the electronic ground state) is small compared to our required energy resolution, such that we can define a Compton-shifted energy variable $E^* = E + \Delta E$ and re-express (4) in approximate form:

$$I_{incoh}(E, 2\theta) = d\Omega_{det} N_{atoms} \frac{d\sigma_t}{d\Omega} \tau_{coh}(E, 2\theta) I_{incident}(E^*) \int_0^\infty dE' \, S_{incoh}(k,\omega). \quad (6)$$

The total scattered intensity, in units of photons/sr, is then

$$I_{total}(E) =$$
$$d\Omega_{det} N_{atoms} \tau(E, 2\theta) \frac{d\sigma_t}{d\Omega} \left[ I_{incident}(E) f(k)^2 S(k) + I_{incident}(E^*) \int_0^\infty dE' \, S_{incoh}(k,\omega) \right]. \quad (7)$$

Note that $\int_0^\infty dE' \, S_{incoh}(k,\omega) = N$ (the atomic number of the scattering species) in the high-$k$ limit of the impulse approximation[51] and takes on smaller values at lower momentum transfers; by comparison, $f(0)^2 = N^2$, and $S(k)$ is of order unity. Therefore the first (coherent) term in $I_{total}(E)$ dominates for heavier elements or for sufficiently small momentum transfers. In Fig. 5 (a) and (b) we compare $f(k)^2$ to the incoherent background scattering for the above-described model. These results lead us to expect that the background in an ED-XRD experiment will not substantially limit the ability to observe the desired coherent scattering.

**II.C. Spectrometers for detection of ED-XRD**

We now consider two different detection options. As shown in the schematic of Fig. 2, one or both of an x-ray CCD and a HOPG-based spectrometer may be used as energy-sensitive detectors. Simulated spectra for both follow in section III. Throughout the remainder of the paper the following experimental parameters are used: $d_{source}$ = 1 cm; target dimensions (for both B and Al): 0.25 mm × 0.25 mm × 0.1 mm; scattering angle $2\theta$ = 135 degrees. These choices will be motivated below.

The modeled CCD has a 2-dimensional square grid of 2200 x 2200 pixels, with a pixel edge length of 13.5 microns. A quantum efficiency of 1 is assumed. The optimal distance between the detector and the target is determined by the competing demands of high signal collection and high rejection of two-photon events on single pixels. We find it reasonable to balance these demands by selecting a single photon-hit regime with an expectation value $p$ of 0.1 photon hits per pixel. At a given scattering angle and in the absence of addition of any special absorbers between the target and the CCD other than a Be filter for low-energy photon rejection, $p$ is determined by the working distance of the CCD and the scattered intensity off the target. The working distance is not a highly-constrained parameter; it must merely be sufficiently large that backgrounds from the high neutron flux and other stray radiations are likely to be substantially suppressed. An upper bound on target intensity arises from signal broadening due to the finite angular size subtended by the target relative to the backlighter. We require this geometrical broadening in momentum transfer, $\Delta k$, to satisfy $\Delta k / k < 0.05$, such that it is sufficiently small compared to the intrinsic scale of structure in $S(k)$. We label the angle subtended by the target $\Delta\theta_t$ and express the geometrical broadening in terms of it: $\Delta k / k = \cot\theta\, \Delta\theta_t$. At $2\theta = 135$ degrees the maximal $\Delta\theta_t$ is approximately 0.1 radians, which corresponds to a sample length of 1 mm.

The task of presenting a modeled HOPG spectrum in a non-configuration-specific manner is complicated by the significant dependence of the spectrometer's energy range on several geometric parameters. Using the labeling of Fig. 6 and, as an example, the spectrometer geometry described by Fig. 7 (a), the differential in $k$ for scattering from the target is

$$dk = \frac{\delta k}{\delta E} dE + \frac{\delta k}{\delta(2\theta)} d(2\theta) = \frac{E}{\hbar c}(\sin(\theta)\cot\theta_B - \cos(\theta))d\theta_B. \qquad (9)$$

As a crystal of length $l$ located a distance $F$ from the target subtends an angle of approximately $(l/F)\sin\theta_B$, we can directly use (9) to calculate the range $\Delta k$ covered by an analyzer crystal as a function of $k$, $2\theta$, and the choice of HOPG reflection. Fig. 7 illustrates this. Salient features of $\Delta k(k, 2\theta)$ are that it is asymptotically linear in $k$ and depends weakly on $2\theta$ everywhere except at low $k$.

That said, we can choose a typical configuration for an HOPG spectrometer and generate a detected spectrum that spans the entire range of $k$ with which we are concerned. Conceptually, this is done by repeatedly rotating the crystal to different central $\theta_B$ to acquire narrow spectra in different ranges of $k$ and then stitching together the resulting spectra. This spectrum, which is henceforth referred to as the "HOPG source spectrum", does not represent a realistic data set, since acquiring it in a single shot would require prohibitively many analyzer crystals, but it does serve as a convenient compilation of the ensemble of possible experimental configurations; the exact choice of spectrometer configuration for a given experiment depends on some prior knowledge of the desired $k$ range, as we discuss below.

The modeled HOPG spectrometer is qualitatively similar to several instruments that have previously been fielded for x-ray Thomson scattering studies at OMEGA [53-55]. For our modeled instrument, the HOPG diffractive element operates on the 002 reflection, has a mosaic spread of 0.3 degrees, is taken to be a flat square with side length $l = 12$ cm, and is located at a distance $F = 25$ cm from the target. The energy-dependent integral reflectivity of the HOPG is based on computed reflectivity curves [56] for an HOPG crystal having a mosaic spread of 0.3 degrees. Denoting $r$ as the peak reflectivity and $\omega$ as the FWHM of the reflectivity curve, the angular integral reflectivity, $\Delta\theta_B$, is approximately $r\omega$; equivalently, the integral reflectivity in energy units is $\Delta E = E \cot\theta_B \Delta\theta_B$. We define $E_{max}$ and $E_{min}$ as the maximum and minimum energies diffracted by the crystal. For isotropically-scattered photons with a fixed energy $E$ between $E_{max}$ and $E_{min}$ the probability of reflection is $\eta = \Omega_0 \Delta E/(4\pi(E_{max} - E_{min}))$, where $\Omega_0 = (l/F)^2 \sin\theta_B$ is the solid angle subtended by the crystal relative to the source (units of sr). Correspondingly, the detected spectrum resulting from $I_{incident}$ on the target is $I_d(E) = 4\pi\eta dI_{incident}(E)/d\Omega = \Delta\theta_B(l/F)dI_{incident}(E)/d\Omega$. For reference, $\Delta\theta_B l/F = 7 \times 10^{-4}$ at 4 keV. It will be seen in the next section that the resulting net collection efficiency in a given achievable energy band is several orders of magnitude higher than that of the CCD.

# III  Results and discussion

There is good reason to believe, heuristically, that the above-described experimental configurations for ED-XRD should determine $S(k)$ with adequate statistics. Numerous past x-ray Thomson scattering experiments at laser plasma facilities have measured the inelastic portion of $S(k,\omega)$ using narrow pulse backlighters for illumination [25, 27, 28, 57-60]. Above 2 keV, a broad-band thermal backlighter has approximately 100 times the photon conversion efficiency of a short-pulse Cu $K_\alpha$ backlighter (Fig. 3); additionally, the elastic scattering cross section is typically larger than the Compton cross section, as discussed in section II and shown in Fig. 5 (a) and 5 (b). Thus, ED-XRD should offer vastly higher signal intensity than (quasi-monochromatic) x-ray Thomson scattering using a metal-foil backlighter, and therefore better statistics.

In Fig. 8 we present $I_d(E)$ defined in section II, filtered by a 20 μm Be foil (to reject low-energy photons) for liquid boron acquired on a CCD alongside the equivalent HOPG source spectrum. The highlighted region of the HOPG source spectrum shows the energy range covered by a specific configuration: a 12-cm long HOPG crystal at distance $F = 25$ cm from the target, oriented such that the detected spectrum is centered on the main correlation peak in $S(k)$. This crystal size results in a solid angle subtended by the crystal similar to that in existing high-efficiency HOPG spectrometers [43, 61]. Figure 9 shows the CCD and HOPG spectra for shock-compressed Al in this same format. $S(k)$ reconstructed for B and Al is presented in Figs. 10 and 11, respectively. In both these figures the $k$-range probed by the specific spectrometer configuration is highlighted. All reconstructed $S(k)$ curves, including those without background subtraction, show a well-defined correlation peak. Note that the uncorrected curves overshoot the experimental $S(k)$ at large $k$; this is a result of the monotonically-increasing Compton background (as well as double-counts, for the CCD). This background decreases relative to the XRD signal for larger atomic numbers, as seen by comparison of Figs. 10 and 11. The HOPG source spectrum exhibits excellent statistics (error bars < 2 percent) relative to the CCD over the entire plotted energy range. While deteriorating at high energy, the CCD spectrum also has good statistics (error bars < 5 percent) below 4.5 keV.

The relative merits of the two detectors are dictated by particular features of the ED-XRD configuration and the spectrum probed. Despite the substantially better energy resolution of an HOPG spectrometer compared to a Fano-noise limited CCD, energy resolution is a poor criterion for comparison: at 135 degrees scattering angle, the $k$-width of features in $S(k)$ corresponds to a width in energy greater than 500 eV, substantially larger than the resolution of both the CCD and the HOPG spectrometer. Instead, the leading limitation on data quality is shot noise at high $k$ due to the sharp decay of the source spectrum intensity with increasing energy.

The latter limitation is severe only for the CCD, (1) because of vastly lower overall counts and especially (2) because the intense low-energy portion of the incident spectrum is 'echoed' as double-counts at higher energy. In simulated CCD spectra the double-count contribution to the detected spectrum outweighed the single-count contribution above 5 keV. This double-count noise cannot be reduced by varying $p$: in the single photon-hit regime, the number of double-hits on single pixels scales as $p^2$; the associated Poisson noise scales as $p$. Single photon counts also scale as $p$; as a result, above 5 keV varying $p$ has little effect on the signal-to-noise (*i.e.* single-to-double-count) ratio. It is instead highly preferable to carry out an ED-XRD experiment in near-backscatter geometry, such that the range of $k$ in which $S(k)$ has structure is probed by a lower-energy region of the backlighter spectrum. In fact, the only means of significantly improving data quality on a single hit CCD are (1) using a detector with more pixels to improve statistics, and (2) moving the detector closer to backscatter.

The two spectrometer types offer a variety of configurations adapted to experimental situations in which different $k$-ranges need to be probed. If the goal is to locate the main correlation peak in $S(k)$, a single-HOPG crystal spectrometer is a viable option (illustrated, as mentioned above, in Figs. 10 and 11). On the other hand, if the scientific goal requires a significantly wider $k$-range, then a CCD detector in single photon counting mode, multiple HOPG analyzer crystals, or both are necessary. Despite the CCD's relatively poor signal to noise ratio, there is substantial motivation for performing ED-XRD using both spectrometer types if the full $k$ range is desired. In such a dual configuration the CCD would provide low-noise data up to approximately 4.5 keV with one or more HOPG spectrometers covering the remainder of the energy spectrum, corresponding to a reduced range of $\theta_B$ from 21 to 49 degrees, for which a modest number of analyzer crystals would be required.

The above results establish single-shot ED-XRD as a viable method for use at OMEGA, even for systems with only liquid-like, isotropic short-range order; this observation clearly extends to fine, isotropic polycrystalline systems where the structure in $S(k)$ can only be sharper. We note that pump laser energy is 30 times larger at NIF than at OMEGA and that the ratio of backlighter fluences exceeds that factor due to the higher backlighter electron temperature at NIF [41]. Consequently, ED-XRD is also viable at NIF where the higher backlighter fluence may allow a substantial reduction in solid angle subtended by an HOPG spectrometer, compared to present calculations. This would in turn allow incorporating a larger number of spectrometers in a single diagnostic module. CCD-based studies at NIF are also, in principle, viable but may run into technical difficulties related to neutron backgrounds or difficulty in shielding from electromagnetic pulses.

## IV. Conclusions

We report a photometric study of the viability of single-shot investigation of the isotropic static structure factor $S(k)$ in experiments using a broadband x-ray backlighter as the source for energy-dispersive x-ray diffraction (ED-XRD). The results are extremely favorable, and indicate that single-shot ED-XRD can be used at OMEGA or NIF. A standard scientific-grade x-ray CCD camera operating in single-photon counting mode suffices for many studies, but exhibits degraded performance at high momentum transfers due to the rapid decrease of incident flux at higher photon energy. On the other hand, a typical HOPG-based wavelength dispersive spectrometer has exceptional count rates in any selected $k$ range, but its limited energy range may require either the use of multiple spectrometers or of a single compound spectrometer having multiple analyzer crystals.

**Acknowledgements**

We thank Brian Mattern, Tilo Doeppner, Philip Nilson, Barukh Yaakobi, Christian Stoeckl, Yuan Ping, Justin Wark, and Andrew Higginbotham for helpful discussions. This work was supported by the US Department of Energy, Office of Science, Fusion Energy Sciences and the National Nuclear Security Administration, through grant DE-SC0008580.

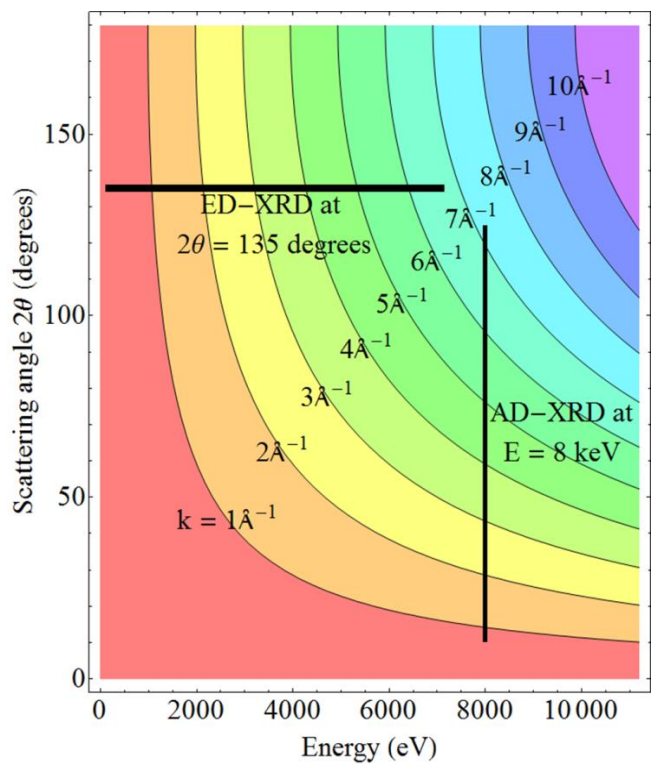

**Figure 1.** Contours of equal momentum transfer $k$ (labeled in units of Å$^{-1}$) in energy and scattering angle. Angle-dispersive x-ray diffraction (AD-XRD) and energy-dispersive x-ray diffraction (ED-XRD) take vertical and horizontal cuts, respectively, to achieve broad coverage in $k$ and thus obtain information about the radial distribution function.

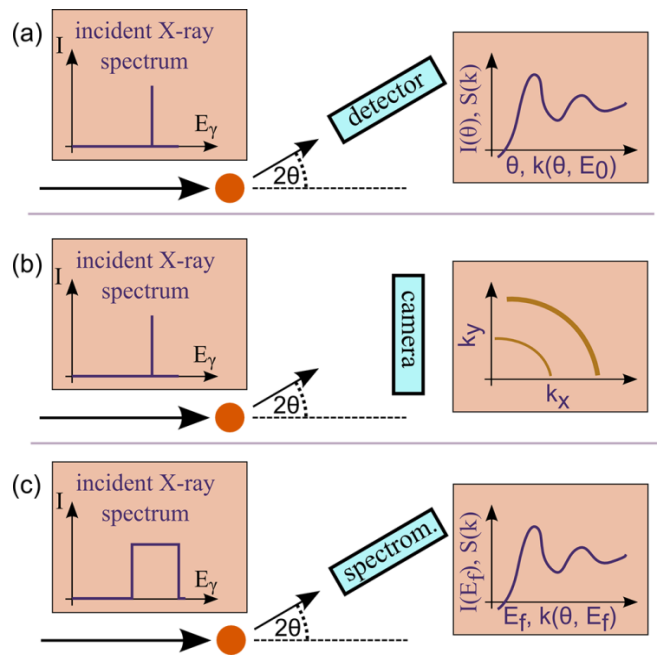

**Figure 2.** Schematic representations of (a) and (b) angle-dispersive x-ray diffraction, compared to (c) energy-dispersive x-ray diffraction (ED-XRD).

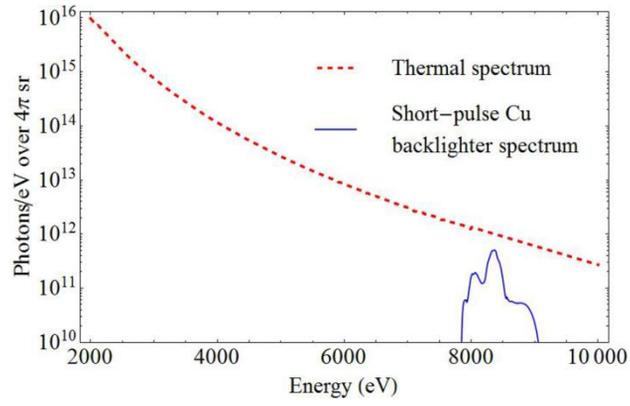

**Figure 3.** Red: experimental thermal backlighter spectrum from OMEGA [43]. Blue: a short-pulse Cu $K_\alpha$ backlighter spectrum, based on scaling of results from a lower energy laser system to a 2.5 kJ, 10 ps laser shot at OMEGA [44, 45].

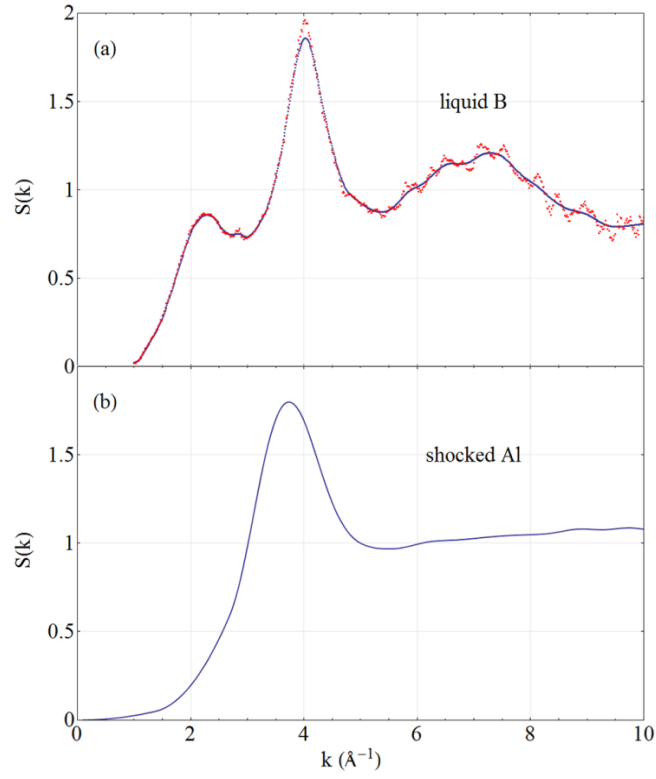

**Figure 4.** (a): Liquid structure factor of B at 2400K. The original data (red) of Krishnan *et al.* [46] contains sharp unphysical noise; we therefore use the filtered interpolation (blue) of the data for $S(k)$ throughout this paper. (b): Equivalent theoretical curve for shock-compressed Al at electron density $n_e = 5.4 \times 10^{23}$ cm$^{-3}$ and temperature $T_e = 10$ eV based on Ma *et al.* [35] The curve is based on an approximate treatment of this system's atomic form factor (see the text for details).

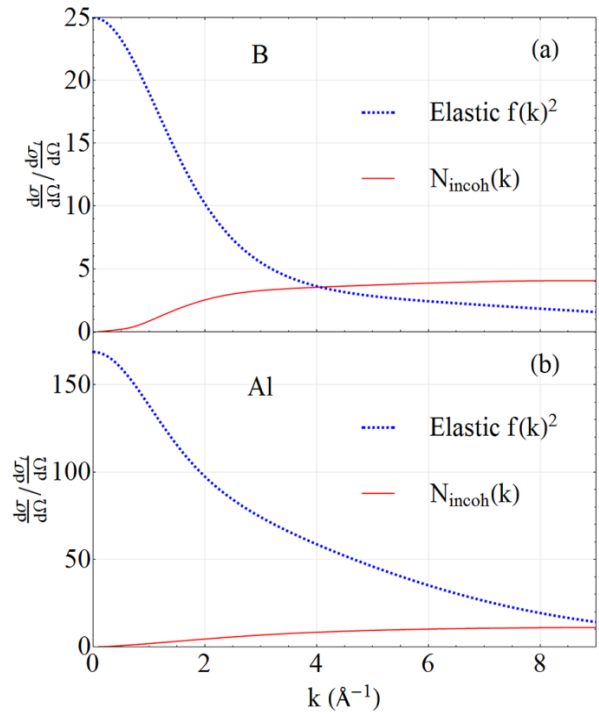

**Figure 5.** Elastic and inelastic contributions to the differential cross section of (a) boron and (b) aluminum. The elastic cross sections are based on tabulated values of $f(k)$. The inelastic differential cross sections, defined by $N_{incoh}(k) = \int_0^\infty dE'\, S(k,\omega')$, are based on $S(k,\omega')$ generated from $f$-summed, truncated Compton profiles in the impulse approximation. See the text for further details.

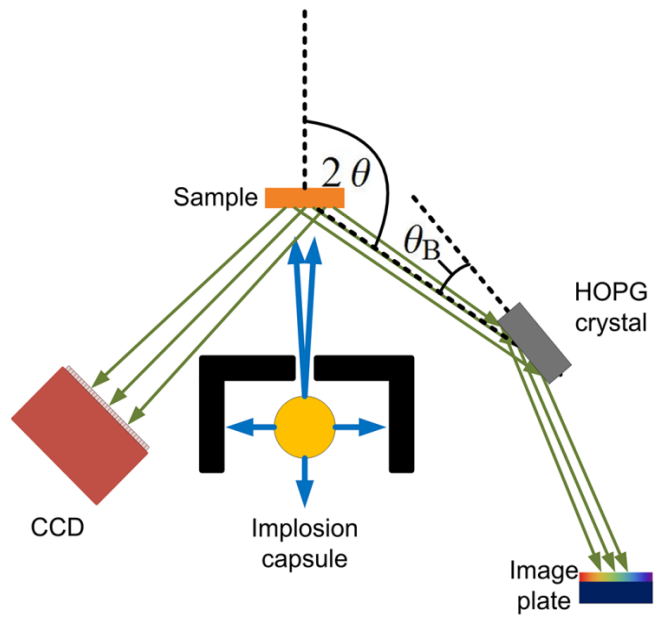

**Figure 6.** Experimental configuration for ED-XRD at a laser shock facility. A long pulse-driven CH capsule emits a broad thermal spectrum. Scattering from the target is observed using an HOPG spectrometer or a CCD in the single-photon hit regime.

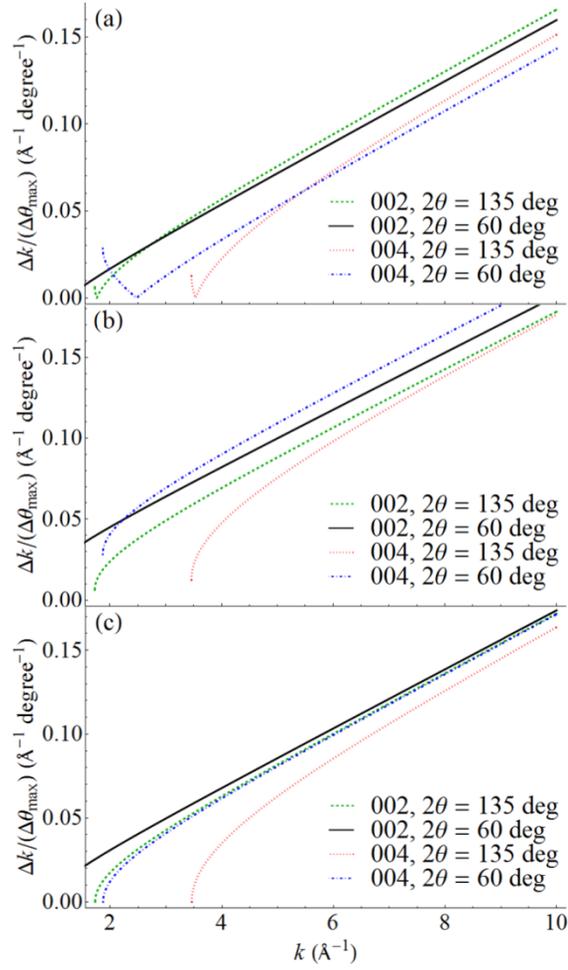

**Figure 7.** Range $\Delta k$ in momentum transfer of scattering off the target probed by a small HOPG crystal per degree of its maximum subtended angle, $\theta_{max}$, for three spectrometer geometries that involve the same position (but different rotations) of the HOPG crystal: (a) the detector located in the target scattering plane and away from the axis passing through the backlighter and target, (b) the detector located in the target scattering plane and near the axis passing through the backlighter and target, and (c) the detector located such that it, the target, and the HOPG crystal define a plane perpendicular to the scattering plane. $\theta_{max}$ denotes the maximum possible subtended angle of the HOPG crystal given a fixed spectrometer working distance $F$; i.e., for a crystal of length $l$, the maximum subtended angle is $\theta_{max} = l/F$.

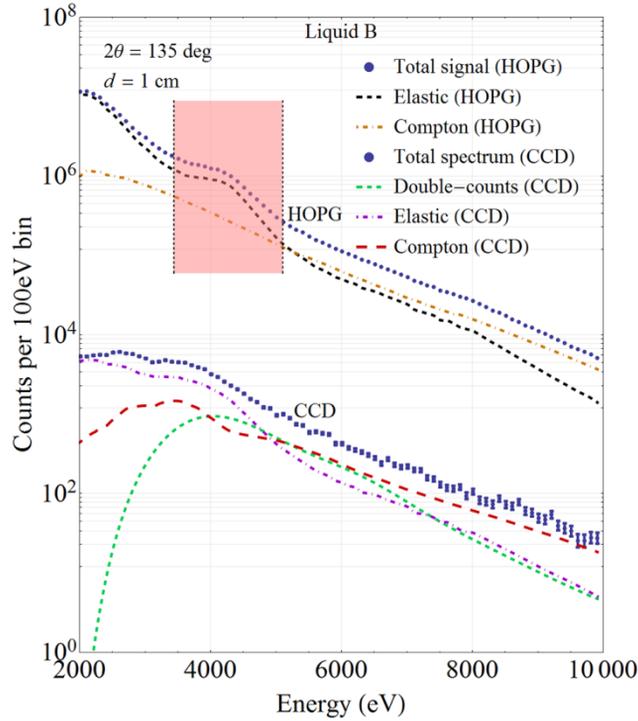

**Figure 8.** Photon-energy histograms for energy-dispersive diffraction spectra of liquid boron on CCD and HOPG spectrometers. The expectation value of photon counts/pixel on the CCD is $p = 0.1$. The energy range of a specific HOPG configuration using a 12-cm long HOPG analyzer is denoted by the shaded region, the width of which corresponds to the spectrometer configuration of Fig. 7 (a). The spectrometer's focal length is 25 cm, and the length of the crystal in the non-energy dispersive orientation is 12 cm; both spectrometers are positioned at $2\theta = 135$ deg. A 20-μm thick Be foil is used to reject low-energy photons. Error bars in the HOPG histogram are smaller than the size of the symbols.

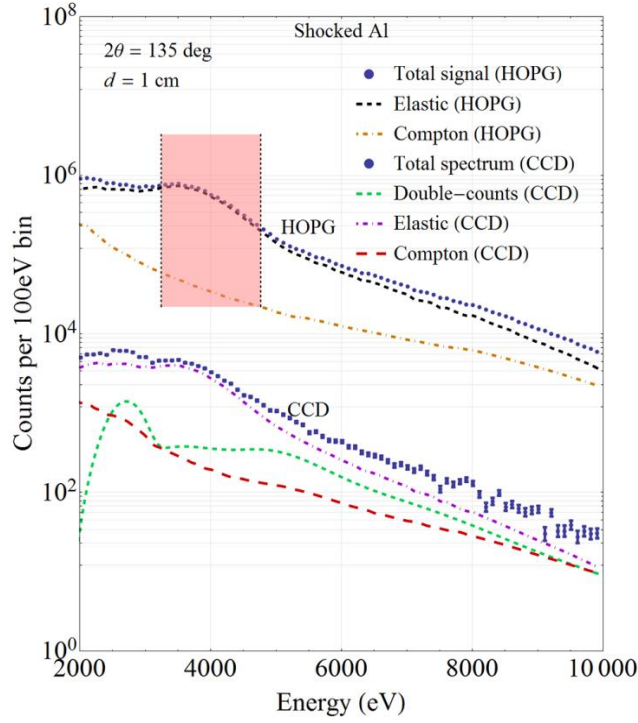

**Figure 9.** Photon-energy histograms on CCD and HOPG spectrometers for shock-compressed Al at electron density $n_e$ = 5.4 × $10^{23}$ cm$^{-3}$ and temperature $T_e$ = 10 eV, using Ma *et al.*'s best-fitting theoretical model to their experimental results for $f(k)^2 \, S(k)$ [35], and assuming $f(k)$ of ambient Al. The expectation value of photon counts/pixel on the CCD is $p$ = 0.1. The energy range of a specific HOPG configuration using a 12-cm long HOPG analyzer is denoted by the shaded region, the width of which corresponds to the spectrometer configuration of Fig. 7 (a). The spectrometer's focal length is 25 cm, and the length of the crystal in the non-energy dispersive orientation is 12 cm; both spectrometers are positioned at $2\theta = 135$ deg. A 20-μm thick Be foil is used to reject low-energy photons. Error bars in the HOPG histogram are smaller than 1% (not shown).

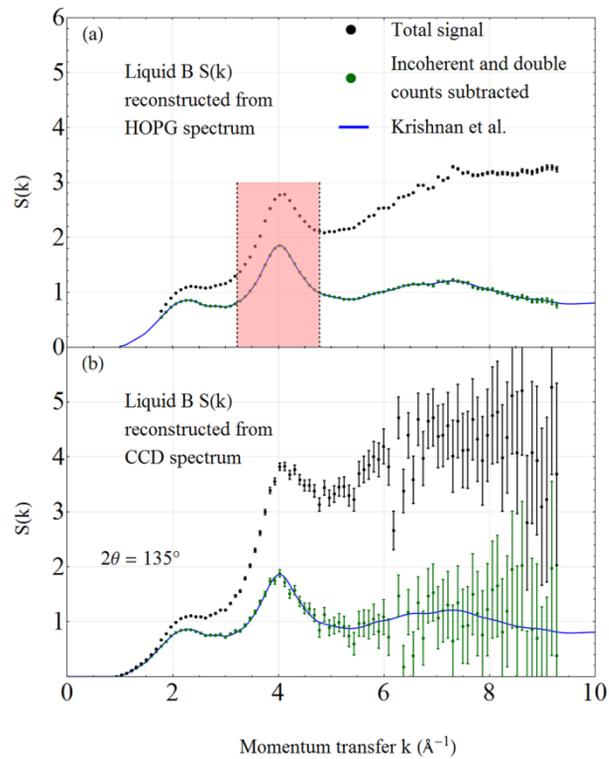

**Figure 10.** $S(k)$ for liquid boron reconstructed from simulated energy-dispersive spectra of Fig. 8 for (a) an HOPG spectrometer and (b) a CCD, with and without subtraction of Compton background and photon double-counts. Data bin size is 100 eV. The shaded $k$-range in the HOPG spectrum corresponds to the spectrometer configuration described in the text, and is centered about the main correlation peak in $S(k)$.

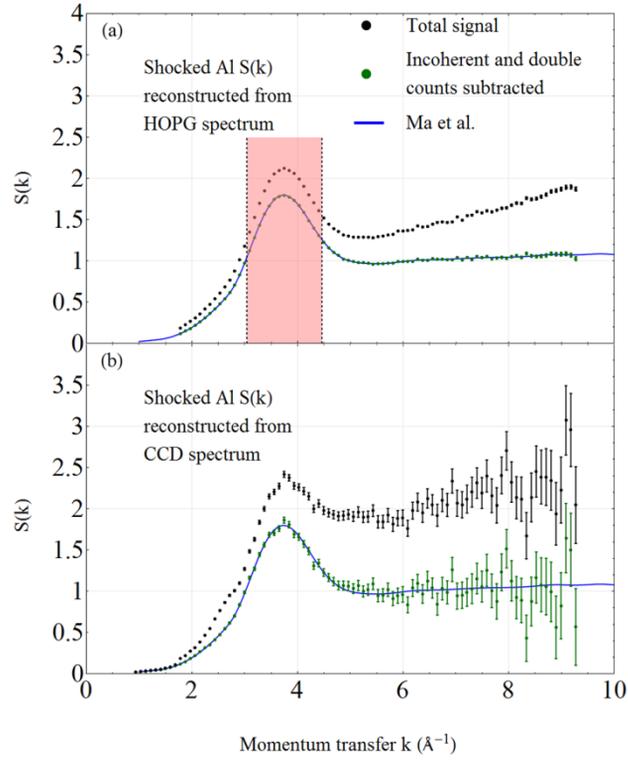

**Figure 11.** In blue: X-ray structure factor $S(k)$ for shock-compressed Al computed from Ma, *et al.* [35]. Overlaid with $S(k)$ reconstructed from the spectra of Fig. 9 for (a) an HOPG spectrometer and (b) a CCD. The data bin size is 100 eV. The shaded $k$-range in the HOPG spectrum corresponds to the spectrometer configuration described in the text and is centered about the main correlation peak in $S(k)$.